\documentclass[twocolumn,prb,showpacs,preprintnumbers,amsmath,amssymb,superscriptaddress]{revtex4-1}

\usepackage{amsmath}
\usepackage{amssymb}
\usepackage{graphicx}
\usepackage[dvipsnames]{xcolor}
\usepackage[colorlinks,linkcolor=blue,citecolor=blue,urlcolor=blue]{hyperref}
\usepackage{stix}
\usepackage{tikz}

\usepackage{ulem}
\usepackage{bm}

\graphicspath{{./figures/}}

\definecolor{lime}{HTML}{A6CE39}
\DeclareRobustCommand{\orcidicon}{
    \begin{tikzpicture}
    \draw[lime, fill=lime] (0,0) 
    circle [radius=0.16] 
    node[white] {{\fontfamily{qag}\selectfont \tiny ID}};
    \draw[white, fill=white] (-0.0625,0.095) 
    circle [radius=0.007];
    \end{tikzpicture}
    \hspace{-2mm}
}

\foreach \x in {A, ..., Z}{\expandafter\xdef\csname orcid\x\endcsname{\noexpand\href{https://orcid.org/\csname orcidauthor\x\endcsname}
            {\noexpand\orcidicon}}
}

\newcommand{\hc}{\hat{c}}

\newcommand{\hb}{\hat{b}}

\newcommand{\hH}{\hat{H}}
\newcommand{\hJ}{\hat{J}}

\newcommand{\hn}{\hat{n}}

\newcommand{\brho}{\boldsymbol{\rho}}

\newcommand{\hbpsi}{\hat{\boldsymbol{\psi}}}

\newcommand{\eqq}[1]{\begin{align} #1 \end{align}}
\newcommand{\bA}{{\bm A}}
\newcommand{\bE}{{\bm E}}

\newcommand{\bP}{{\bm P}}

\newcommand{\bU}{{\bf U}}
\newcommand{\br}{{\bm r}}
\newcommand{\bk}{{\bm k}}
\newcommand{\beps}{\boldsymbol{\epsilon}}
\newcommand{\bd}{{\bf d}}
\newcommand{\hbJ}{\hat{\bm J}}
\newcommand{\hbP}{\hat{\bm P}}
\newcommand{\bh}{{\bm h}}

\begin{document}
\title{Doping and gap-size dependence of high-harmonic generation in graphene : \\Importance of consistent formulation of light-matter coupling }

\author{Yuta Murakami\orcidA{}}
\email{yuta.murakami@phys.titech.ac.jp}
\affiliation{Department of Physics, Tokyo Institute of Technology, Meguro, Tokyo 152-8551, Japan}
\affiliation{Center for Emergent Matter Science, RIKEN, Wako, Saitama 351-0198, Japan}
\author{Michael Sch\"{u}ler\orcidB{}}
\affiliation{Condensed Matter Theory Group, Paul Scherrer Institute, CH-5232 Villigen PSI, Switzerland}
\date{\today}

\begin{abstract}
High-harmonic generation (HHG) in solids is a fundamental nonlinear phenomenon, which can be efficiently controlled by modifying system parameters such as doping-level and temperature.
In order to correctly predict the dependence of HHG on these parameters, consistent theoretical formulation of the light-matter coupling is crucial. 
Recently, contributions to the current that are often missing in the HHG analysis based on the semiconductor Bloch equations have been pointed out [J. Wilhelm, et.al. PRB {\bf 103} 125419 (2021)].
In this paper, by systematically analyzing the doping and gap-size dependence of HHG in gapped graphene, we discuss the practical impact of such terms.
In particular, we focus on the role of the current $J_{\rm ra}^{(2)}$, which originates from the change of the intraband dipole via interband transition.
When the gap is small and the system is close to half filling, intraband and interband currents mostly cancel, thus suppressing the HHG signal -- an important property that is broken when neglecting $J_{\rm ra}^{(2)}$. Furthermore, without $J_{\rm ra}^{(2)}$, the doping and gap-size dependence of HHG becomes qualitatively different from the full evaluation.
Our results demonstrate the importance of the consistent expression of the current to study the parameter dependence of HHG for the small gap systems.
\end{abstract}

\maketitle

\section{Introduction}	
Recent development of laser technology in the terahertz and infrared regime enables the study of various nonlinear phenomena in condensed matters originating from strong light-matter coupling~\cite{Kruchinin2018}.
Important examples include dielectric breakdown~\cite{Yamakawa2017}, Bloch oscillations~\cite{Ghimire2011NatPhys} as well as Floquet engineering~\cite{McIver2020,Oka2019review}.
The high-harmonic generation (HHG) is also a fundamental example of such nonlinear phenomena~\cite{Corkum2007,Krausz2009RMP,Ghimire2019}.
While HHG was originally observed and studied in gas systems~\cite{Ferray_1988}, its scope has been recently extended to condensed matters, in particular semiconductors and semimetals~\cite{Ghimire2011NatPhys,Schubert2014,Luu2015,Vampa2015Nature,Langer2016Nature,Hohenleutner2015Nature,Ndabashimiye2016,Liu2017,You2016,Yoshikawa2017Science,Hafez2018,Kaneshima2018,Yoshikawa2019,Matsunaga2020PRL,Sanari2020PRB,Schmid2021}.
One important aspect of condensed matters is the sensitivity of material properties against system parameters such as doping-level and temperature.
This feature opens the interesting possibility of controlling HHG in condensed matters using active degrees of freedoms~\cite{Ikeda2020PRR,Nishidome2020,Uchida2022PRL,Tamaya2021PRB}.
For example, strong doping dependence of the HHG spectrum has been reported in carbon nanotubes, where the doping level is controlled by gating~\cite{Nishidome2020}. 

To explore the intriguing possibility of controlling HHG, consistent understanding of the origin of HHG is essential. 
There are several approaches to theoretically study HHG in solids~\cite{Otobe2016,Ikemachi2017,Tancogne-Dejean2017b,Tancogne-Dejean2017,Golde2008,Higuchi2014,Vampa2014PRL,Vampa2015PRB,Wu2015,Luu2016,Hansen2017,Osika2017,Chacon2020PRB,Ikeda2018PRA,Tamaya2016,Floss2018,Silva2019PRB,Markus2020SOC,Yue2020PRA,Taya2021PRB}.
One major strategy is the time dependent density functional theory (TDDFT)~\cite{Otobe2016,Hansen2017,Ikemachi2017,Tancogne-Dejean2017b,Tancogne-Dejean2017,Floss2018}.
In principle, TDDFT can provide an ab-initio way to study HHG. However, its accuracy is limited by the inevitable approximations to the exchange-correlation functional and relaxation effects.
Another major approach complementary to TDDFT is to study model systems with several bands around the Fermi level applying the semiconductor Bloch equations (SBEs)~\cite{Golde2008,Higuchi2014,Vampa2014PRL,Vampa2015PRB,Luu2016,Osika2017,Chacon2020PRB,Tamaya2016,Silva2019PRB,Markus2020SOC}.
The SBEs are formulated based on the single-particle density matrix (SPDM), and allows us to disentangle contributions to HHG and to easily introduce the relaxation and dephasing effects at least phenomenologically.
These approaches revealed that many features of HHG in semiconductors and semimetals can be explained as the dynamics of independent electrons (independent particle picture).
Furthermore, it has been pointed out that there are two major contributions to HHG  in semiconductors: the intraband and interband currents.
The former essentially represents the intraband acceleration of electrons (holes) in the conduction (valence) band, while the latter represents the change of the interband polarization.
The dominant contribution depends on systems, excitation conditions and the order of harmonics, and the two contributions may cancel with each other in some occasions.
Still, the separation of contributions is helpful to obtain the physical picture of the HHG mechanism in solids.
For example, HHG from the interband current can be understood by the so-called three step model~\cite{Vampa2014PRL,Vampa2015PRB,Ikemachi2017}.

Despite the success, there still remains ambiguity in the treatment based on the SBEs originating from the choice of gauges of the light and bases for electronic states~\cite{Wilhelm2021PRB,Yue2022tutorial}.
Different works in the literature often use different representations and different classification of contributions, and thus consistency between these studies is not fully clear.
Recently, Wilhelm {\it et.al.} rederived the SBEs and clarified the relation between different representations~\cite{Wilhelm2021PRB}. 
They point out the existence of two types of currents that are often neglected in the HHG analyses based on the SBEs:
(i) The contribution to the current originating from the change of the intraband dipole via interband transition. We call it $J_{\rm ra}^{(2)}$ in this paper.
This term contributes to the intraband current, when the intraband current is defined as the derivative of the intraband dipole moment.
(ii) The contribution originating from the dephasing term phenomenologically introduced to the SBEs.
In Ref.~\onlinecite{Wilhelm2021PRB}, the authors demonstrate the importance of these contributions using the massless Dirac model with a fast dephasing time.
Still, in order to fully understand the role of these often-neglected terms and their practical impact, further systematic analyses are necessary.

In this paper, we study the doping and gap-size dependence of HHG in gapped graphene and reveal the role of often-neglected contributions, in particular, the role of $J_{\rm ra}^{(2)}$.
 We show that, when the gap is small and the system is close to half filling, the cancellation between the intraband and interband currents is severe 
 and the inclusion of the contribution of $J_{\rm ra}^{(2)}$ to HHG becomes important. 
 On the other hand, when the gap becomes large compared to the excitation frequency, the contribution from the interband current becomes dominant and the effect of $J_{\rm ra}^{(2)}$ becomes relatively marginal.
 We demonstrate that, without $J_{\rm ra}^{(2)}$, the massless or non-doped system is predicted to be favorable for the large HHG intensity, while, in the full evaluation, the HHG intensity shows nonmonotonic behavior against the gap-size and the doping level. Our results demonstrate the importance of the consistent expression of the current to correctly predict the parameter dependence of HHG for the small gap systems.
These insights should be also relevant for HHG from the surface states of topological insulators ~\cite{Schmid2021,Baykusheva2021PRA,Baykusheva2021}.

This paper is organized as follows. 
In Sec.~\ref{sec:general}, focusing on the two-band model, we revisit the formulation of the light-matter coupling and clarify the relation between different representations.
In Sec.~\ref{sec:graphene}, we introduce the tight-binding model for gapped graphene applying the general formulation discussed in Sec.~\ref{sec:general}.
We also introduce the effective Dirac models.
In Sec.~\ref{sec:results}, we present the numerical results, examine the doping and gap dependence of HHG and discuss the role of different components of the currents to HHG.
The summary is given in the last section.

\section{Formulation: General statements} \label{sec:general}
In this section, we revisit the formulation of the light-matter problem based on the SBEs and clarify the relation between frequently-used representations to be self-contained.
We note that a general discussion is already given in Refs.~\onlinecite{Wilhelm2021PRB} and \onlinecite{Yue2022tutorial}.
For simplicity, we focus on a specific problem, i.e. the tight-binding model consisting of two well-localized Wannier states per unit cell.
Our setup is directly applicable to graphene and hexagonal boron nitride (hBN). The extension to mutiorbital cases is straightforward, which is relevant for the transition metal dichalcogenides such as  WSe$_2$ and MoS$_2$~\cite{Liu2013PRB,Fang2015PRB}. 
In the following, we use the dipole approximation (neglecting the spatial dependence of the field). We also assume $\langle \psi_i | \mathbf{\hat{r}}| \psi_j\rangle = \mathbf{r}_i \delta_{ij}$, where $\mathbf{\hat{r}}$  is the position operator and $| \psi_i\rangle$ is a well-localized Wannier state centered at $\mathbf{r}_i$. Namely, the dipole matrix element between the different Wannier states is zero.
 Because of  this assumption, the light-matter coupling in the dipole gauge is equivalent to the Peierls substitution.
As our starting point, we employ the length gauge.
 In this gauge, the Hamiltonian for the light-matter coupled problem is  expressed as 
 \eqq{
  \hH^{\rm L}(t) = - \sum_{i\neq j} t_{{\rm hop},i j}  \;  \hc^\dagger_{i} \hc_{j}  + \sum_i V_i \hn_i - q\sum_i {\bm E}(t)\cdot{\bm r}_i \hn_i, \label{eq:H_LG_space}
 }
where $\hc^\dagger_i$ is the creation operator of an electron at the $i$th site, corresponding to the Wannier state $|\psi_i\rangle$, and $\hn_i = \hc_i^\dagger \hc_i$.
$t_{{\rm hop},ij}$ is the transfer integral from the $j$th site to the $i$th site, $V_i$ sets the energy level, $q$ is the charge of the electron, ${\bm E}(t)$ is the electric field and ${\bm r}_i$ is the position vector of the $i$th site.
We omit the spin index assuming the paramagnetic phase.
The Hamiltonian~\eqref{eq:H_LG_space} is the low-energy tight-binding model, which is obtained by the projection of the first-principles Hamiltonian in the length gauge to the space spanned by the specified Wannier states~\cite{Michael2021PRB}.  Another way to construct the low-energy model is to start from the minimal coupling, i.e. projection from the velocity gauge. 
Although first-principles Hamiltonians in the length gauge and the velocity gauge are equivalent for finite systems (which can in principle be infinitely big), they are not equivalent after the projection.
As discussed in Ref.~\onlinecite{Michael2021PRB}, the main difference of the optical response originates from the inequivalent treatment of the diamagnetic current. The high-energy response such as HHG is expected to be less sensitive against the choice of the projected models.

\begin{figure}[t]
  \centering
    \hspace{-0.cm}
    \vspace{0.0cm}
\includegraphics[width=85mm]{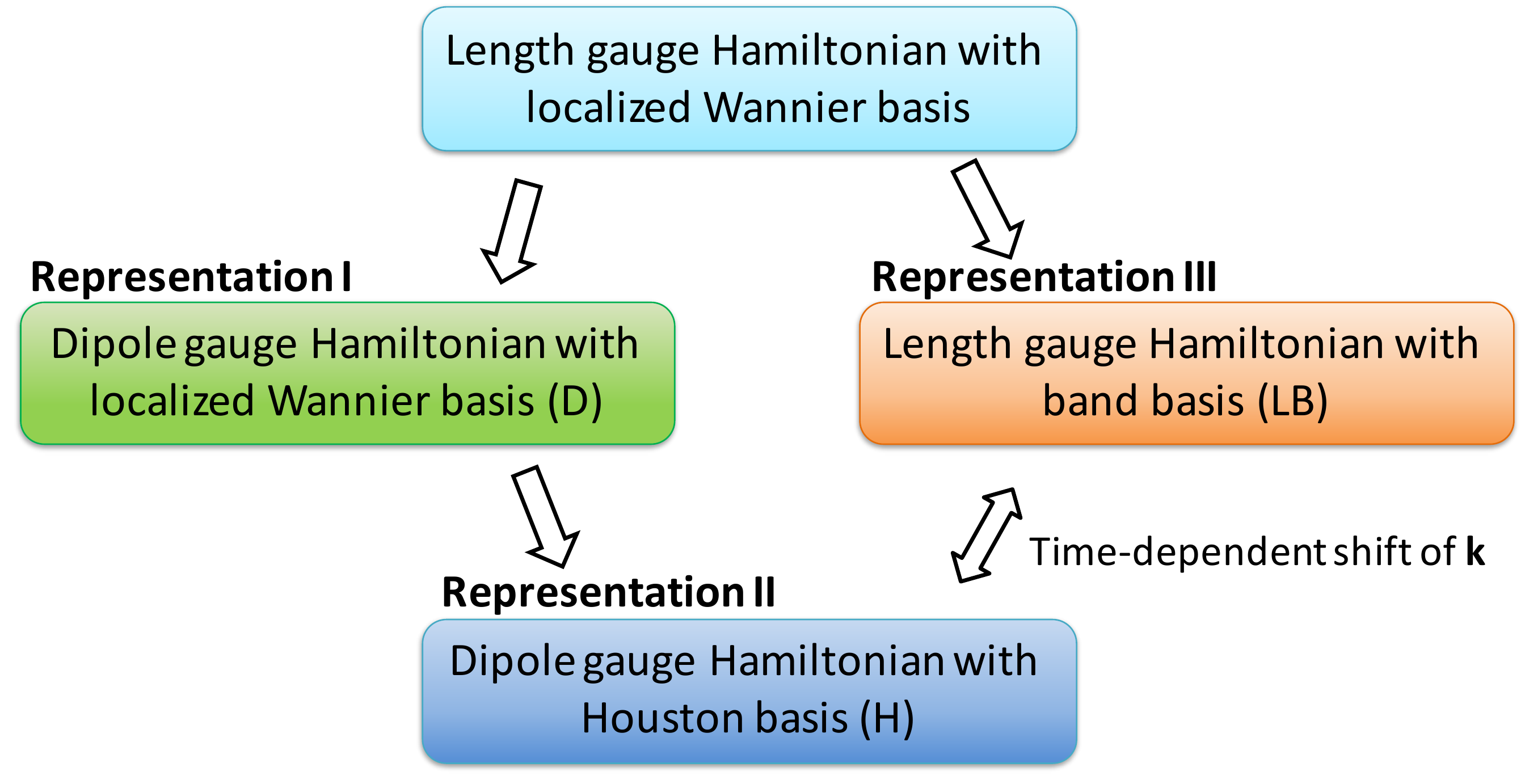} 
  \caption{Summary of the relation between different representations discussed in this paper.}
  \label{fig:summary_representation}
\end{figure}

In the following, we will demonstrate how frequently used representations are obtained via unitary transformations from Eq.~\eqref{eq:H_LG_space}, see Fig.~\ref{fig:summary_representation}.
 In all representations, the Hamiltonian ($ \hH(t)$) is quadratic. In order to study the time-evolution of the system, we focus on SPDM,
 \eqq{
 \rho_{ij}(t) = \langle \hc^\dagger_j(t) \hc_i(t) \rangle.
 }
 Here, $\langle \cdots  \rangle$ is the expectation value with the grand canonical ensemble, and $\hc^\dagger (t)$ indicates the Heisenberg representation of $\hc^\dagger$.
 Introducing the matrix elements of the Hamiltonian as $h_{ij}(t) = \langle {\rm vac}| \hc_i \hH(t) \hc^\dagger_j | {\rm vac}\rangle$, the time evolution of SPDM (the von Neumann equation or the quantum master equation) is  expressed as 
\eqq{
\partial_t \brho(t) = i[\brho(t), {\bm h}(t) ] + \partial_t \brho(t)|_{\rm corr}.
}
Here, $\brho(t) ({\bm h}(t))$ expresses the matrix with elements  $\rho_{ij}(t)$ ($h_{ij}(t)$). We set $\hbar$ unity in this paper.
The last term indicates the contribution from the correlations originating from electron-electron interactions, electron-phonon interactions and impurities, 
although these are absent in our Hamiltonian~\eqref{eq:H_LG_space}.  
$\partial_t \brho(t)|_{\rm corr}$ can be directly evaluated by explicitly including these terms in the Hamiltonian and using the diagrammatic expansions~\cite{Aoki2013,stefanucci_nonequilibrium_2013,Nessi2020,ridley2022manybody}.
However, the direct microscopic evaluation of $ \partial_t \brho(t)|_{\rm corr}$ is computationally expensive, and instead the correlation effects are often taken into account phenomenologically through the relaxation time approximation
[see Sec.~\ref{sec:relaxation}].
 
 The intensity of HHG is evaluated from the current $J(t)$ induced by the external field as $I_{\rm HHG}(\omega)=|\omega J(\omega)|^2$, where $J(\omega)$ is the Fourier transform of $J(t)$.
$J(t)$ can be directly evaluated as the expectation value of the current operator $\hat{J}$ or from the time derivative of the expectation value of the polarization $\hat{P}$ using $\brho(t)$.
Both approaches yield the identical result if the time evolution with respect to the Hamiltonian is solved exactly.

\subsection{Representation I: Dipole gauge expressed with localized Wannier basis}

Applying the unitary transformation $\hc^\dagger_i \rightarrow e^{iq {\bf A}(t)\cdot {\bm r}_i}\hc^\dagger_i$ to the Hamiltonian \eqref{eq:H_LG_space},  we obtain the dipole gauge Hamiltonian~\cite{Michael2021PRB}
 \eqq{
  \hH^{\rm D}(t) = - \sum_{i\neq j} t_{{\rm hop},i j} e^{iq{\bf A}(t)\cdot {\bm r}_{ij}}  \;  \hc^\dagger_{i} \hc_{j}  + \sum_i V_i \hn_i, \label{eq:H_DG_space}
 }
 where ${\bm r}_{ij} = {\bm r}_i - {\bm r}_j$ and ${\bf A}(t)$ is the vector potential. The latter is related with the electric field as ${\bm E}(t) = -\partial_t {\bf A}(t)$.
In this gauge, the light-matter coupling is taken into account through the Peierls phase.

It is natural to express this Hamiltonian in the momentum space representation applying the periodic boundary condition to the Hamiltonian~\eqref{eq:H_DG_space}
and using the Bloch states defined by the Wannier states as $|\psi_{\bf k,\alpha}\rangle = \frac{1}{\sqrt{N}} \sum_{i\in\alpha}e^{i{\bm k}\cdot{\bm r}_i} |\psi_{i}\rangle$.
Namely, we introduce the creation operators as 
$ \hc^\dagger_{{\bm k}\alpha } = \frac{1}{\sqrt{N}} \sum_{i\in \alpha } e^{i{\bm k}\cdot {\bm r}_i}  \hc^\dagger_{i}, \label{eq:c_kA} $
where $\alpha =$ A,B indicates the sublattices. Here, $N$ is the number of unit cells in the system.
The resulting expression is
\eqq{
\hH^{\rm D}(t) = \sum_\bk 
\begin{bmatrix}
\hc^\dagger_{\bk A} & \hc^\dagger_{\bk B}
\end{bmatrix}
{\bm h}(\bk-q\bA(t))
\begin{bmatrix}
\hc_{\bk A}\\
\hc_{\bk B}
\end{bmatrix}, \label{eq:H_DG_wannier}
}
where ${\bm h}(\bk)$ is obtained by the Fourier transformation of $t_{{\rm hop},i j}$ in terms of $\br_{ij}$. Note that ${\bm h}(\bk)$ is in general not diagonal.
In the following, we express $\bk-q\bA(t)$ as $\bk(t)$ and $\hbpsi^\dagger_\bk = [\hc^\dagger_{\bk A},  \hc^\dagger_{\bk B}]$.

In this representation, the von Neumann equation for SPDM, $\rho^{\rm D}_{\alpha\beta,\bk}(t) = \langle \hc^{\dagger}_{\bk\beta}(t) \hc_{\bk\alpha}(t) \rangle$,
is expressed as 
\eqq{
\partial_t \brho^{\rm D}_{\bk}(t) = i[\brho^{\rm D}_{\bk}(t), {\bm h}(\bk(t))]. \label{eq:vonNeumann_DW}
}
The operator of the current of the $a$-direction is defined as $\hat{J}_a(t)=-\delta H^{\rm D}(t) / \delta A_a(t)$.
More explicitly, it is expressed as  $\hat{J}_a(t) = \sum_\bk \hat{J}_{\bk a}(t)$ with 
\eqq{
\hat{J}_{\bk a}(t) =  q\hbpsi^\dagger_\bk [\partial_{a} {\bm h}(\bk(t))] \hbpsi_\bk,
}
where we defined $\partial_a = \partial / \partial {k_a}$.

Since the expression of ${\bm h}(\bk)$ can be easily evaluated, this representation is an obvious choice for the numerical implementation.
However, for classifying different contributions to HHG or when including phenomenological relaxation terms, it is more convenient to choose the basis set that diagonalizes  ${\bm h}(\bk)$~\cite{Sato2021PRB}.
For the following change of representation, we assume that the system is gapped (no degeneracy of the eigenvalue of ${\bm h}(\bk)$ at each $\bk$).

\subsection{Representation II: Dipole gauge expressed with the Houston basis }
Now we consider the representation using the instantaneous eigenstates of the time-dependent Hamiltonian $\hH_\bk(t)(\equiv \hbpsi^\dagger_\bk {\bm h}(\bk(t)) \hbpsi_\bk)$, i.e. the Houston basis~\cite{Wu2015,Wilhelm2021PRB,Yue2022tutorial}. 
The representation is obtained by the time-dependent unitary transformation of Eq.~\eqref{eq:H_DG_wannier} with $\hat{\mathcal{U}}^\dagger(t) = \prod_\bk \hat{\mathcal{U}}^\dagger_\bk(t)$
, where $\hat{\mathcal{U}}^\dagger_\bk(t)$ satisfies
\eqq{
\hat{\mathcal{U}}^\dagger_\bk(t)
\begin{bmatrix}
\hc_{\bk A}\\
\hc_{\bk B}
\end{bmatrix} 
\hat{\mathcal{U}}_\bk(t)
= \bU(\bk(t))
\begin{bmatrix}
\hc_{\bk A}\\
\hc_{\bk B}
\end{bmatrix}.
}
Here $\bU(\bk)^\dagger {\bm h}(\bk) \bU(\bk) = \beps(\bk)$ and $\beps(\bk) = {\rm diag} [\epsilon_0(\bk), \epsilon_1(\bk) ]$ is a diagonal matrix.
We note that the choice of $\bU(\bk)$ is not unique and there exists a gauge freedom.
After this transformation, the meaning of $\hc^\dagger_{\bk \alpha}$ is changed and it represents the instantaneous eigenstate of $\hH_\bk(t)$ in the original representation.
In order to clarify the difference of the meaning, we express $\hc^{\dagger}_{\bk A}$ ($\hc^{\dagger}_{\bk B}$)  as $\hb^{\dagger}_{\bk 0}$ ($\hb^{\dagger}_{\bk 1}$) after this transformation in the following,
and introduce $\hbpsi^{'\dagger}_{\bk} = [\hb^{\dagger}_{\bk 0}, \hb^{\dagger}_{\bk 1}]$.
The resultant Hamiltonian ($\hH^{\rm H}_{\bk} (t) = \hat{\mathcal{U}}^\dagger_\bk(t) \hH^{\rm D}_{\bk}(t) \hat{\mathcal{U}}_\bk(t)  + i(\partial_t \hat{\mathcal{U}}^\dagger_\bk(t) ) \hat{\mathcal{U}}_\bk(t)$) becomes 
\eqq{
\hH^{\rm H}_{\bk} (t) =
\hbpsi^{'\dagger}_\bk
{\beps}(\bk(t))
\hbpsi^{'}_\bk
-q\sum_a E_a(t) \hat{\bf d}_a({\bm k}(t)), 
}
where $\hat{\bf d}_a({\bm k}(t)) = \hbpsi^{'\dagger}_\bk {\bf d}_a(\bk(t)) \hbpsi^{'}_\bk$ is the dipole moment for the direction $a$, and ${\bf d}_a(\bk) = i \bU^\dagger (\bk)[\partial_{a} \bU(\bk)]$ is the (non-abelian) Berry connection, which plays the role of dipole matrix elements. 
Note that the Hamiltonian in this representation is now diagonalized at each $\bk$ and time, at least in the adiabatic limit ($\Omega \rightarrow 0$ and ${\bm E}(t)\simeq {\bf 0}$).
After the transformation the expression of the current ($\hbJ'_{\bk}(t) =  \hat{\mathcal{U}}^\dagger_\bk(t) \hbJ_{\bk}(t) \hat{\mathcal{U}}_\bk(t) $) becomes 
\eqq{
\hat{J}'_{\bk a}(t) =  q\hbpsi^{'\dagger}_\bk  [\partial_{a} \beps(\bk(t))] \hbpsi^{'}_\bk - iq\hbpsi^{'\dagger}_\bk  [ {\bf d}_a(\bk(t)),\beps(\bk(t))] \hbpsi^{'}_\bk. \label{eq:dipole_current}
}
The first term consists of the diagonal components of $\hb^\dagger_{\bk n} \hb_{\bk m}$, i.e. $n=m$, while the second term consists of the off-diagonal components.
In the literature, the first and second terms are sometimes referred to as the intraband and interband currents, respectively~\cite{Ishikawa2010PRB,Bowlan2014PRB,Dimitrovski2017PRB,Candong2018PRA,Mrudul2021PRB}.
However, they are different from the intraband and interband currents defined in terms of the length gauge~\cite{Sipe2000PRB,Kaneko2021PRL} [see Sec.~\ref{sec:rep3}].
For example, the first term only depends on the dispersion of the band, while the intraband current includes the anomalous velocity originating from the topological nature of the wave functions.

The corresponding von Neumann equation for $\rho^{\rm H}_{mn,\bk}(t) = \langle \hb^{\dagger}_{\bk n}(t) \hb_{\bk m}(t) \rangle$
becomes 
\eqq{
\partial_t \brho^{\rm H}_{\bk}(t) = i[\brho^{\rm H}_{\bk}(t), \beps(\bk(t)) -q\sum_a E_a(t) \bd_a(\bk(t)) ]. \label{eq:vN_II}
}
Actually, this representation is closely related to the representation III discussed in the following section.


\subsection{Representation III : Length gauge expressed with band basis} \label{sec:rep3}
Now we come back to the length gauge and express the Hamiltonian \eqref{eq:H_LG_space} using the band basis,
\eqq{
\begin{bmatrix}
\hc_{\bk 0}^\dagger & \hc_{\bk 1}^\dagger
\end{bmatrix} 
= 
\begin{bmatrix}
\hc_{\bk A}^\dagger  & \hc_{\bk B}^\dagger
\end{bmatrix}
\bU(\bk).
}
In this representation, $\hH^{\rm L}(t)$ is expressed as 
\eqq{
\hH^{\rm L}(t) = \sum_\bk 
\begin{bmatrix}
\hc^\dagger_{\bk 0} &\hc^\dagger_{\bk 1} 
\end{bmatrix}
\beps(\bk)
\begin{bmatrix}
\hc_{\bk0} \\\hc_{\bk1}
\end{bmatrix}
-\bE(t) \cdot \hat{\bP}.
}
The important issue is the expression of $\hat{\bP}$, which includes the position operator  [see Eq.~\eqref{eq:H_LG_space}].
Note that the band basis implicitly assumes the periodic boundary condition,  while the position operator is not well-defined in this condition.
As is discussed in Refs.~\onlinecite{Blount1962,Sipe2000PRB}, in the thermodynamic limit, the polarization operator can be interpreted as 
\begin{subequations}\label{eq:polarization}
\eqq{
\hat{\bm P} & = \hat{\bm P}_{\rm ra} + \hat{\bm P}_{\rm er}, \\ \nonumber
 \hat{\bm P}_{\rm ra} &= \hat{\bm P}^{(I)}_{\rm ra} +  \hat{\bm P}^{(II)}_{\rm ra} \nonumber \\
 &= 
  q\sum_\bk 
\begin{bmatrix}
\hc^\dagger_{\bk 0} &\hc^\dagger_{\bk 1} 
\end{bmatrix}
\begin{bmatrix}
\bd_{00}(\bk) & 0\\
0  & \bd_{11}(\bk)
\end{bmatrix}
\begin{bmatrix}
\hc_{\bk 0} \\  \hc_{\bk 1}
\end{bmatrix} \nonumber \\
&+q\sum_{\bk,\bk'}\sum_n \Bigl[i\nabla_\bk \delta(\bk-\bk')\Bigl] \hc^\dagger_{\bk n} \hc_{\bk' n}  ,
\\
 \hat{\bm P}_{\rm er} &= q
  \sum_\bk 
\begin{bmatrix}
\hc^\dagger_{\bk 0} &\hc^\dagger_{\bk 1} 
\end{bmatrix}
\begin{bmatrix}
0 & \bd_{01}(\bk) \\
\bd_{10}(\bk) & 0
\end{bmatrix}
\begin{bmatrix}
\hc_{\bk 0} \\  \hc_{\bk 1}
\end{bmatrix}.}
\end{subequations}
Here, $\boldsymbol{\epsilon}(\bk)$ and $\bd(\bk)$ are the same as in the representation II.
$\hat{\bm P}_{\rm ra}$  is the intraband polarization, which is expressed with diagonal components of $\hc^\dagger_n\hc_m$.
On the other hand, $ \hat{\bm P}_{\rm er}$  is the interband polarization, which is expressed with off-diagonal components of $\hc^\dagger_n\hc_m$.
The current corresponds to the change of the polarization, $\hbJ(t) = -i[\hbP,\hH^{\rm L}(t)]$. 
One can consider two types of currents originating from the intraband and interband polarizations~\cite{Sipe2000PRB,Kaneko2021PRL};
\eqq{
\hbJ_{\rm ra}(t) = -i[\hbP_{\rm ra}(t),\hH^{\rm L}(t)] , \;\;\; \hbJ_{\rm er}(t) = -i[\hbP_{\rm er}(t),\hH^{\rm L}(t)].
}
The explicit expression of the total current along the $a$ axis is
\eqq{
\hJ_a &= 
\sum_\bk
\begin{bmatrix}
\hc^\dagger_{\bk 0} &\hc^\dagger_{\bk 1} 
\end{bmatrix}
\partial_{a}
\beps(\bk)
\begin{bmatrix}
\hc_{\bk 0} \\  \hc_{\bk 1}
\end{bmatrix}  \\ \nonumber
& -i 
\sum_\bk
\begin{bmatrix}
\hc^\dagger_{\bk 0} &\hc^\dagger_{\bk 1} 
\end{bmatrix}
[\bd_a(\bk), \boldsymbol{\epsilon}(\bk)]
\begin{bmatrix}
\hc_{\bk 0} \\  \hc_{\bk 1}
\end{bmatrix}.
} 

The expression of the intraband current becomes $\hJ_{{\rm ra},a}(t)= \hJ^{(1)}_{{\rm ra},a}(t) + \hJ^{(2)}_{{\rm ra},a}(t)$, where
\eqq{
\hJ^{(1)}_{{\rm ra},a}(t)  &= \sum_\bk \sum_n v_{n,a}(\bk,t) \hc^\dagger_{\bk n}  \hc_{\bk n}, \label{eq:J_ra_1} \\
\hJ^{(2)}_{{\rm ra},a}(t)  &= -\bE(t)\sum_{\bk,n\neq m}  {\bm r}_{nm,a}(\bk) \hc^\dagger_{\bk n}  \hc_{\bk m}, 
}
with 
\eqq{
v_{n,a} (\bk,t) & = \partial_a \epsilon_n(\bk) - \Bigl[\bE(t) \times (\nabla_\bk \times  \bd_{nn}(\bk))\Bigl]_a \label{eq:velocity} \\
 {\bm r}_{nm,a}(\bk)& = \partial_{a} \bd_{nm} (\bk) -i(d_{nn,a}(\bk)-d_{mm,a}(\bk)){\bf d}_{nm}(\bk). 
}
$\hJ^{(1)}_{{\rm ra},a}$ consists of the diagonal terms in terms of $\hc^\dagger_n\hc_m$.
 $\nabla_\bk \times  \bd_{nn}(\bk)$ represents the Berry curvature, and the second term in Eq.~\eqref{eq:velocity} is the anomalous velocity. 
On the other hand, $\hJ^{(2)}_{{\rm ra},a}$  consists of the {\it off-diagonal} terms.
$\hJ^{(2)}_{{\rm ra},a}$ originates from $-i[\hbP_{\rm ra},-\bE(t) \cdot \hat{\bP}_{\rm er}] $.
Physically, this indicates the change of the intraband polarization by the interband excitation via $-\bE(t) \cdot \hat{\bP}_{\rm er}$.
We note that this term corresponds to the shift current~\cite{Sipe2000PRB,Morimoto2016,Kaneko2021PRL,Souza2022}. 

The interband current is expressed as 
\eqq{
& \hJ_{{\rm er},a} 
 = -i 
\sum_\bk
\begin{bmatrix}
\hc^\dagger_{\bk 0} &\hc^\dagger_{\bk 1} 
\end{bmatrix}
[\bd_a(\bk), \boldsymbol{\epsilon}(\bk)]
\begin{bmatrix}
\hc_{\bk 0} \\  \hc_{\bk 1}
\end{bmatrix}   \nonumber\\
&\quad -\bE(t)\sum_{\bk,n, m}  \tilde{\bm r}_{nm,a}(\bk) \hc^\dagger_{\bk n}  \hc_{\bk m},  \\
&\tilde{\bm r}_{nm,a}  \nonumber\\
& =
\begin{cases}
-i (d_{n\bar{n},a}(\bk)\bd_{\bar{n}n}(\bk)- d_{\bar{n}n,a}(\bk)\bd_{n\bar{n}}(\bk))\;\;\;(n= m)\\
-\nabla_\bk d_{nm,a}(\bk) +id_{nm,a}(\bk)(\bd_{nn}(\bk)-\bd_{mm}(\bk)) \;\;\;(n\neq m).
\end{cases}
}
Here $\bar{n} = 1-n$.
Note that our definition of the intraband and interband currents are based on the types of the polarization as Ref.~\onlinecite{Sipe2000PRB}.
On the other hand, the authors of Ref.~\onlinecite{Wilhelm2021PRB} define $\hJ^{(1)}_{{\rm ra},a}$ as the intraband current and all remaining terms is the interband contribution.
We also note that $\hat{\bm J}_{\rm er}$ includes a term $-i[\hbP_{\rm er},-\bE(t) \cdot \hat{\bP}_{\rm ra}] (\equiv \hat{\bm J}^{(2)}_{\rm er})$, which resembles $\hat{\bm J}^{(2)}_{\rm ra}$.
Indeed, if we focus on the linearly polarized filed and the current along the field direction, we have $\hat{J}^{(2)}_{\rm er}=-\hat{J}^{(2)}_{\rm ra}$.

The corresponding von Neumann equation for  $\rho^{\rm LB}_{mn,\bk}(t) = \langle \hc^\dagger_{\bk n}(t) \hc_{\bk m}(t) \rangle$ is
\eqq{\partial_t \brho^{\rm LB}_{\bk} (t) &= i [\brho^{\rm LB}_{\bk}(t) , {\bm h}^{\rm LB}(\bk,t)] - (\bE(t)\cdot\nabla_\bk) \brho^{\rm LB}_{\bk} (t)\label{eq:SBE_ususal}} 
with  ${\bm h}^{\rm LB}(\bk,t) = \beps(\bk) -\sum_a E_a(t)\bd_a(\bk)$.
This form of SBEs has been often used for the analysis of HHG~\cite{Vampa2014PRL,Vampa2015PRB,Luu2016,Chacon2020PRB}.
Furthermore, if we define $\tilde{\brho}^{\rm LB}_{\bk}(t)\equiv \brho^{\rm LB}_{\bk-q\bA(t)} (t)$,
we have 
\eqq{
\partial_t  \tilde{\brho}^{\rm LB}_{\bk} (t) &= i [\tilde{\brho}^{\rm LB}_{\bk}(t) , {\bm h}^{\rm LB}(\bk-q\bA(t),t)].
}
This equation is the same as Eq.~\eqref{eq:vN_II} for the SPDM in the representation II.
Since the initial SPDM is the same between the representation II and the representation III, $\brho^{\rm H}_{\bk}(t)= \tilde{\brho}^{\rm LB}_{\bk}(t)$.
This also justifies the expression of the polarization \eqref{eq:polarization}.

In the analysis of HHG based on the SBEs in the form of Eq.~\eqref{eq:SBE_ususal}, the intraband and interband currents are evaluated separately~\cite{Vampa2014PRL,Vampa2015PRB,Luu2016,Chacon2020PRB}.
For the intraband current, only the contribution from $\hJ^{(1)}_{{\rm ra},a}$ is often taken into account, and it is important to clarify the role of $\hJ^{(2)}_{{\rm ra},a}$.
On the other hand, the interband current is often evaluated through a derivative of the expectation value of $\hat{\bm P}_{\rm er}$.

\subsection{Phenomenological relaxation and dephasing} \label{sec:relaxation}
In real materials, relaxation and dephasing of excited carriers occur due to the electron-electron interactions, electron-phonon interactions and disorders. 
These effects are often taken into account via  phenomenological terms in the von Neumann equation.
They are usually introduced for the relaxation process of the occupation of the bands and the dephasing process between the bands.
In the representation II, the phenomenological von Neumann equation becomes
\eqq{
\partial_t \brho^{\rm H}_{\bk}(t) &= i[\brho^{\rm H}_{\bk}(t), \beps(\bk(t)) -\sum_a E_a(t) \bd_a(\bk(t)) ] \nonumber \\
& - \frac{\brho^{\rm H}_{{\rm diag},\bk}(t) - \brho^{\rm H}_{{\rm eq},\bk(t)}}{T_1} - \frac{\brho^{\rm H}_{{\rm off},\bk}(t)}{T_2}. \label{eq:von_Neumann_Houston2}
}
Here, $\brho_{{\rm diag}}$ indicates a matrix consisting of diagonal components of $\brho$, while $\brho_{{\rm off}}$ indicates a matrix consisting of off-diagonal components of $\brho$.
The second term represents the relaxation process, where the occupation (the diagonal terms of $\brho^{\rm H}$) approaches the equilibrium value with a time scale $T_1$.
The third term expresses the dephasing process, where the off-diagonal components of $\brho^{\rm H}$ approaches zero (the equilibrium value) with a time scale $T_2$.
The corresponding expression in the representation III is naturally obtained from Eq.~\eqref{eq:von_Neumann_Houston2}.
In the representation I, since $ \brho^{\rm DW}_\bk(t)  =  {\bf U}(\bk(t)) \brho^{\rm H}_\bk(t) {\bf U}^\dagger(\bk(t))$,
 Eq.~\eqref{eq:von_Neumann_Houston2} corresponds to 
\eqq{
 \partial_t \brho^{\rm DW}_\bk(t) &= i[\brho^{\rm DW}_\bk(t),\bh(\bk(t))]  - \frac{\brho^{\rm DW}_\bk(t) - \brho^{\rm DW}_{eq,\bk(t)}}{T_1} \nonumber\\
& + \Bigl(\frac{1}{T_1}-\frac{1}{T_2}\Bigl) {\bf U}(\bk(t))\brho^{\rm H}_{{\rm off},\bk(t)}{\bf U}^\dagger(\bk(t)). \label{eq:vN_eq_T1T2_DW}
}

Upon introducing phenomenological relaxation and dephasing terms, attention needs to be payed to the following issue.
The current obtained directly from evaluating the expression for $\hat{\bm J}$, i.e. $\langle \hat{\bm J}\rangle$, and from the derivative of the expectation value of the polarization, i.e. $\partial_t \langle \hat{\bm P} \rangle$, are not equivalent any more (without the phenomenological terms they are equivalent). 
We note that this discrepancy corresponds to the current induced by the dephasing, which is pointed out in Ref.~\onlinecite{Wilhelm2021PRB} and is also often neglected in the HHG analysis using SBE~\cite{Vampa2014PRL,Vampa2015PRB}.
Therefore, when one uses small $T_1$ and $T_2$, this subtlety of how to evaluate a certain quantity becomes a practical problem.

\subsection{Lessons}\label{sec:lesson}
From the above section, one can identify the following issues; i) there is a often-neglected term $J_{\rm ra}^{(2)}$ in analyses of 
HHG based on the representation III, and ii) the phenomenological damping term may bring some inconsistency between different ways to evaluate the current~\cite{Wilhelm2021PRB,Yue2022tutorial}.
In the following, we focus on gapped graphene as an example, and discuss how these points are relevant for the doping and gap-size dependence of the HHG spectrum.

 \begin{figure}[t]
  \centering
    \hspace{-0.cm}
    \vspace{0.0cm}
\includegraphics[width=45mm]{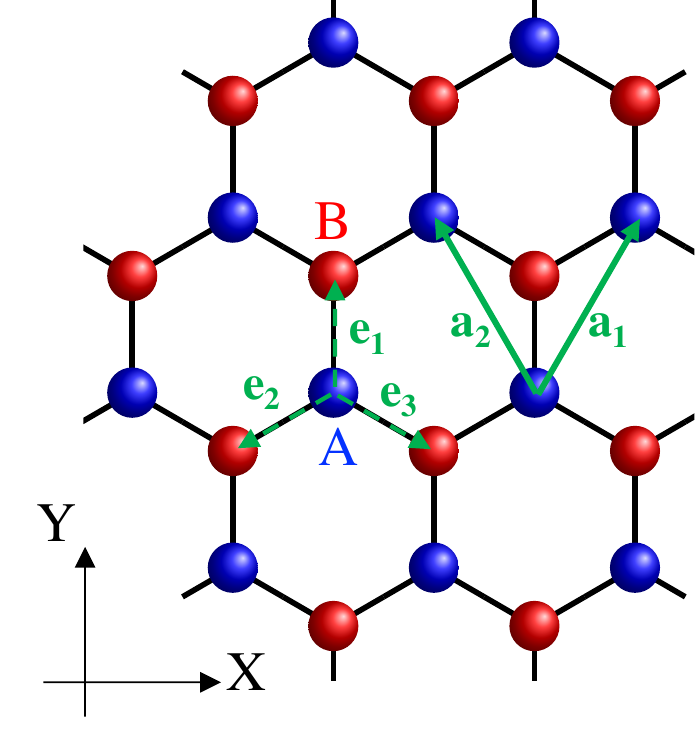} 
  \caption{Tight-binding model on the two-dimensional honeycomb lattice. Blue circles indicate the A sublattice, while red circles indicate the B sublattice.  }
  \label{fig:Graphene}
\end{figure}

\section{Graphene Models} \label{sec:graphene}
In this section, we apply the general formulation discussed in Sec.~\ref{sec:general} to the tight-binding models for gapped graphene.
We note that the same model is also applicable for hBN. 
We consider the two dimensional honeycomb lattice as in Fig.~\ref{fig:Graphene}. We set the length of the bond to unity.

In equilibrium, the tight-binding model is expressed as 
\eqq{
\hH = -t_{\rm hop}\sum_{\langle ij\rangle} \hc^\dagger_{i} \hc_{j} +m\sum_i (-1)^i \hn_i -\mu \sum_i \hn_i.
}
 $\langle ij\rangle$ indicates a pair of the neighboring sites ($\langle ij\rangle \neq \langle ji\rangle$).
$t_{\rm hop}$ is the transfer integral, $m$ is the energy level difference between the A and B sublattices, $(-1)^i = 1$ for $i\in A$, $(-1)^i = -1$ for $i\in B$, and $\mu$ is the chemical potential.


\subsection{Light-matter coupling in representation I}\label{sec:graphene_A}
Assuming that the Wannier state is well localized at the $i$th site, we apply the general formulation in Sec.~\ref{sec:general}.
The Hamiltonian corresponding to Eq.~\eqref{eq:H_DG_space} is
\eqq{
\hH(t) = -t_{\rm hop} \sum_{\langle ij\rangle} e^{iq{\bf A}(t)\cdot {\bm r}_{ij}} \hc^\dagger_{i} \hc_{j} +m\sum_i (-1)^i \hn_i -\mu \sum_i \hn_i.
}
The corresponding current operator is 
$\hat{\bm J}(t) = iqt_{\rm hop} \sum_{\langle ij\rangle} {\bm r}_{ij} e^{iq{\bf A}(t)\cdot {\bm r}_{ij}} \hc^\dagger_{i} \hc_{j}.$

The Hamiltonian corresponding to Eq.~\eqref{eq:H_DG_wannier} is 
\eqq{
\hH(t) &= -t_{\rm hop} \sum_{\bm k} 
\hbpsi^\dagger_{\bm k}
\begin{bmatrix}
0  & F( {\bm k}-q{\bf A}(t)) \\
 F^*( {\bm k}-q{\bf A}(t)) & 0
\end{bmatrix}
\hbpsi_{\bm k} \nonumber
\\& \;\;\;\; + 
 \sum_{\bm k} 
\hbpsi^\dagger_{\bm k}
\begin{bmatrix}
m-\mu  & 0 \\
0 & -m-\mu
\end{bmatrix}
\hbpsi_{\bm k}
,  \label{eq:Graphene_TB_k}
}
while the current is 
\eqq{
\hat{\bm J}(t) =  t_{\rm hop}\sum_a  i{\bf e}_a \sum_{{\bm k}} 
\hbpsi^\dagger_{\bm k}
\begin{bmatrix}
0 &- e^{i({\bm k}-q{\bf A})\cdot {\bf e}_a} \\
 e^{-i({\bm k}-q{\bf A})\cdot {\bf e}_a} & 0
\end{bmatrix}
\hbpsi_{\bm k}.
} 
Here, $F( {\bm k})  =e^{i{\bm k}\cdot {\bf e}_1} + e^{i{\bm k}\cdot {\bf e}_2} + e^{i{\bm k}\cdot {\bf e}_3}$ and ${\bf e}_i$ indicates vectors from a site of the A sublattice to the neighboring sites [see Fig.~\ref{fig:Graphene}].
We note that the mass term yields the band gap of $2m$.

\subsection{Effective Dirac models}\label{sec:dirac}
The above tight-binding model hosts two Dirac points in the Brillouin zone at ${\bm K}$ and ${\bm K'}$, where $F( {\bm K})=0$ and $F( {\bm K}')=0$.
One can focus on the dynamics of electron arounds the Dirac points, 
when the Fermi-level is close to the Dirac points ($\mu$ is not far from $0$), the excitation frequency is small compared to the bandwidth, and the field is not too strong.
The dynamics of electrons can be described by the effective Dirac models, which are obtained by expanding  the  tight-binding model \eqref{eq:Graphene_TB_k} around the Dirac points.

Around ${\bm k}\simeq {\bm K}$ (${\bm k}\simeq {\bm K}'$), we introduce $\delta {\bm k}\equiv {\bm k} - {\bm K}$ ($\delta {\bm k}\equiv {\bm k} - {\bm K}'$ ), expand Eq.~\eqref{eq:Graphene_TB_k} in terms of $\delta {\bm k}(t) \equiv \delta {\bm k}-q{\bf A}(t)$ and regard $e^{-i{\bm K}\cdot {\bf e}_1}\hc^\dagger_{{\bm k}B}$ ($e^{-i{\bm K}'\cdot {\bf e}_1}\hc^\dagger_{{\bm k}B}$ ) as new $\hc^\dagger_{{\bm k}B}$. Finally, we obtain the effective Hamiltonian for each ${\bm k}$:
\begin{widetext}
\eqq{
\hH^{({\bm K},{\bm K}')}_{\bm k} (t) = 
\hbpsi^\dagger_{\bm k}
\begin{bmatrix}
-\mu +m &-\frac{3 t_{\rm hop}}{2} [\pm \delta k_x (t) + i \delta k_y(t)] \\
-\frac{3 t_{\rm hop}}{2} [\pm \delta k_x (t) - i \delta k_y(t)] & -\mu-m
\end{bmatrix}
\hbpsi_{\bm k}, \label{eq:Dirac_1}
}
while the corresponding current operator becomes 
\eqq{
\hat{\bf j}^{({\bm K}, {\bm K}')}_{\bm k} = \mp\frac{3t_{\rm hop}}{2} {\bf e}_x \;
\hbpsi^\dagger_{\bm k}
\begin{bmatrix}
0 &1 \\
1 & 0
\end{bmatrix}
\hbpsi_{\bm k}
-\frac{3t_{\rm hop}}{2} {\bf e}_y \;
\hbpsi^\dagger_{\bm k}
\begin{bmatrix}
0 &i \\
-i & 0
\end{bmatrix}
\hbpsi_{\bm k}.
}
\end{widetext}

\subsection{Implementation}
Using the explicit form of the Hamiltonians shown in Secs.~\ref{sec:graphene_A} and \ref{sec:dirac},
we implement the code based on the representation I, i.e. Eq.~\eqref{eq:vN_eq_T1T2_DW}, for the original tight-binding model and its effective Dirac models.
The different types of currents defined in the representation III are obtained by taking account of the relation between these representations as discussed in Sec.~\ref{sec:rep3}.
A mored detailed explanation on the implementation is found in Appendix.~\ref{sec:Z}.

 \begin{figure*}[tb]
  \centering
    \hspace{-0.cm}
    \vspace{0.0cm}
\includegraphics[width=180mm]{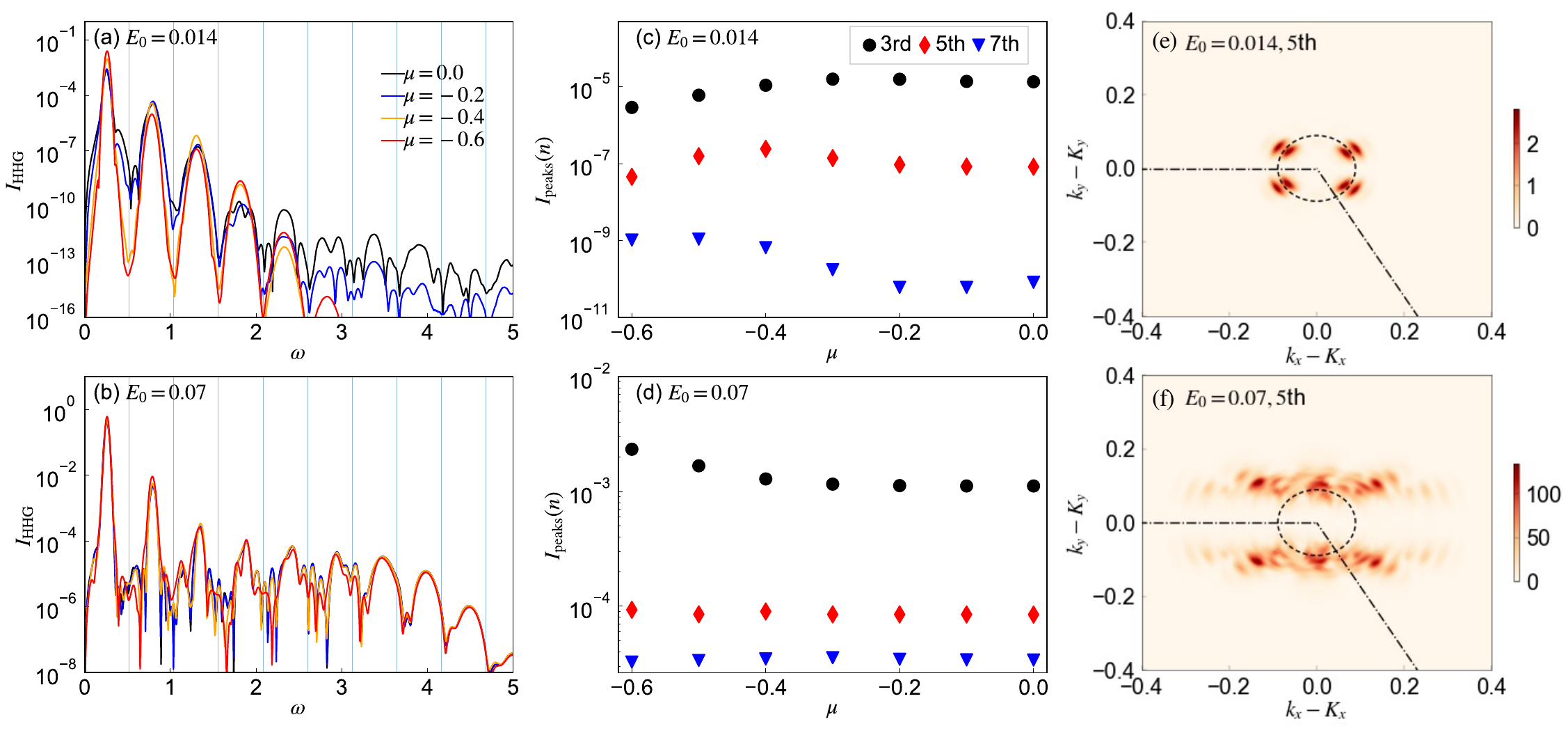} 
  \caption{(a)(b) HHG spectra $I_{\rm HHG}=|\omega J_X(\omega)|^2$ of gapped graphene for indicated values of chemical potential and field strength.  (c)(d) The intensity of the peaks of the HHG spectra $I_{\rm peaks}(n)$ as a function of chemical potential. 
(e)(f) The intensity of the 5th harmonic peak of $I_{\rm HHG,\bk}(\omega) =|\omega J_{\bk,X} (\omega)|^2$ for indicated values of field strength around the Dirac point (${\bm K}=(K_x,K_y)$).
   The dashed circles indicates the Fermi surface for $\mu=-0.4$, while
   the dot-dashed lines indicate the edge of the Brillouin zone of the graphene.
In all cases, we set $t_{\rm hop}=3$, $m=0.001$, $T_1=150$ and $T_2=30$. 
  The parameters of the electric field are $t_0=280,\sigma=40$ and $\Omega=0.26$. These results are obtained from the analysis of the Dirac models. 
  }
  \label{fig:Dirac_HHG_mu_dep}
\end{figure*}

 \begin{figure*}[tb]
  \centering
    \hspace{-0.cm}
    \vspace{0.0cm}
\includegraphics[width=180mm]{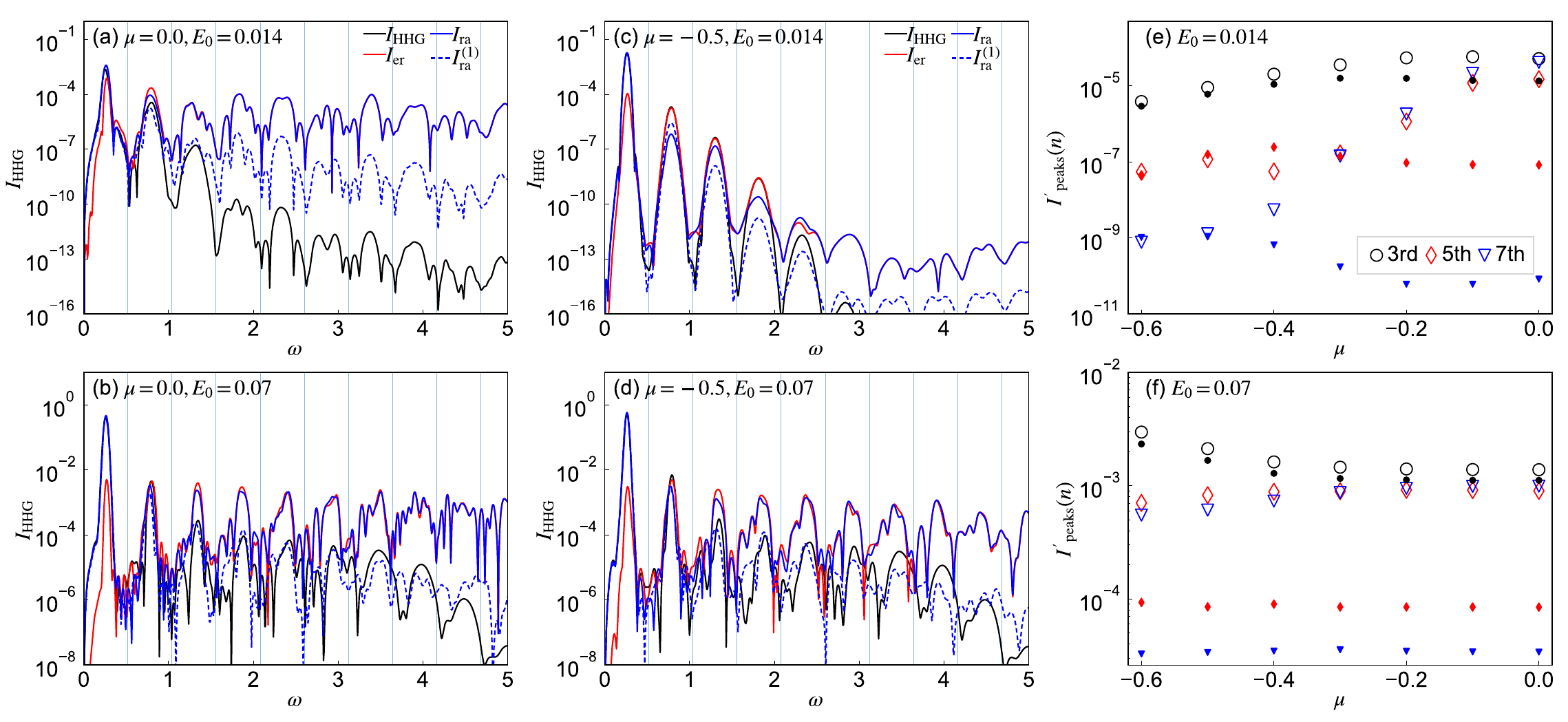} 
  \caption{(a-d) HHG spectra  of gapped graphene for indicated values of chemical potential and field strength.  We compare the contributions from different types of currents; $I_{\rm HHG}=|\omega J_X(\omega)|^2$, $I_{\rm er}=|\omega J_{{\rm er},X}(\omega)|^2$, $I_{\rm ra}=|\omega J_{{\rm ra},X}(\omega)|^2$, $I^{(1)}_{\rm ra}=|\omega J^{(1)}_{{\rm ra},X}(\omega)|^2$.
   (e)(f) The intensity of the peaks of the HHG spectra $I'_{\rm peaks}(n)$, which is  evaluated from $I'_{\rm HHG}(\omega) =|\omega (J_{{\rm er},X}(\omega) + J_{{\rm ra},X}^{(1)}(\omega))|^2$ (i.e. without $J_{\rm ra}^{(2)}$ ), are shown with open markers, as a function of  chemical potential. The filled markers indicates the results from $I_{\rm HHG}(\omega)$ shown in Figs.~\ref{fig:Dirac_HHG_mu_dep}(e)(f).
In all cases, we use $t_{\rm hop}=3$, $m=0.001$, $T_1=150$ and $T_2=30$. The parameters of the electric field are $t_0=280,\sigma=40$ and $\Omega=0.26$.  These results are obtained from the analysis of the Dirac models. }
  \label{fig:Dirac_HHG_detail}
\end{figure*}

\section{Results}\label{sec:results}
In this section, we show the results of the HHG spectra of gapped graphene and discuss the issues  raised in Sec.~\ref{sec:lesson}.
In the following, we set $t_{\rm hop}=3$ and the excitation frequency $\Omega=0.26$. 
Since the hopping of the graphene is roughly $3$ eV, our energy unit corresponds to $1$ eV.
Under this correspondence, the excitation frequency corresponds to $\Omega=0.26$ eV, which is in the mid-infrared regime,
and our time unit approximately corresponds to $0.66$ fs.
This set of parameters is motivated by experiments on graphene and carbon nanotubes~\cite{Yoshikawa2017Science,Nishidome2020}. 
In addition, we set the bond length ($0.246\times \frac{1}{\sqrt{3}}$ nm for graphene) as our unit of length and set the charge $q$ to unity.
With this choice, the field strength of $1$ MV/cm corresponds approximately to $E_0=0.014$ in theory units.

We set the temperature as $T=0.03$, which corresponds to the room temperature.
As for the dephasing time, we set $T_2=30$, which is almost $20$ fs as is recently reported~\cite{Heide2021}.
We set $T_1=150$, which is much larger than $T_2$ as in Ref.~\onlinecite{Sato2021PRB}.
These time scales are reasonable to describe dephasing and relaxation originating from genuine many-body effects.
We note that these time scales are much longer than the time ($T_2=1$ fs) used in Ref.~\onlinecite{Wilhelm2021PRB}.
Such short dephasing times of a few fs have been often used in previous studies. 
As pointed out in Refs.~\onlinecite{Floss2018,Kilen2020PRL}, it can be regarded as a crude way to mimic the dephasing by the propagation of light and the inhomogeneity of the field strength.

In the following, we mainly show the results obtained from the analysis of the effective Dirac model,
since the expression of the dipole moment ${\bf d}$ is much simpler in this model compared to the original graphene model [see Appendix~\ref{sec:A}].
We have checked that the full HHG spectrum $I_{\rm HHG}$ obtained from the Dirac model and the original graphene model agrees reasonably well
for the excitation conditions considered here [see Appendix~\ref{sec:B}].

\subsection{Linearly polarized light: Doping dependence}

 \begin{figure*}[tb]
  \centering
    \hspace{-0.cm}
    \vspace{0.0cm}
\includegraphics[width=180mm]{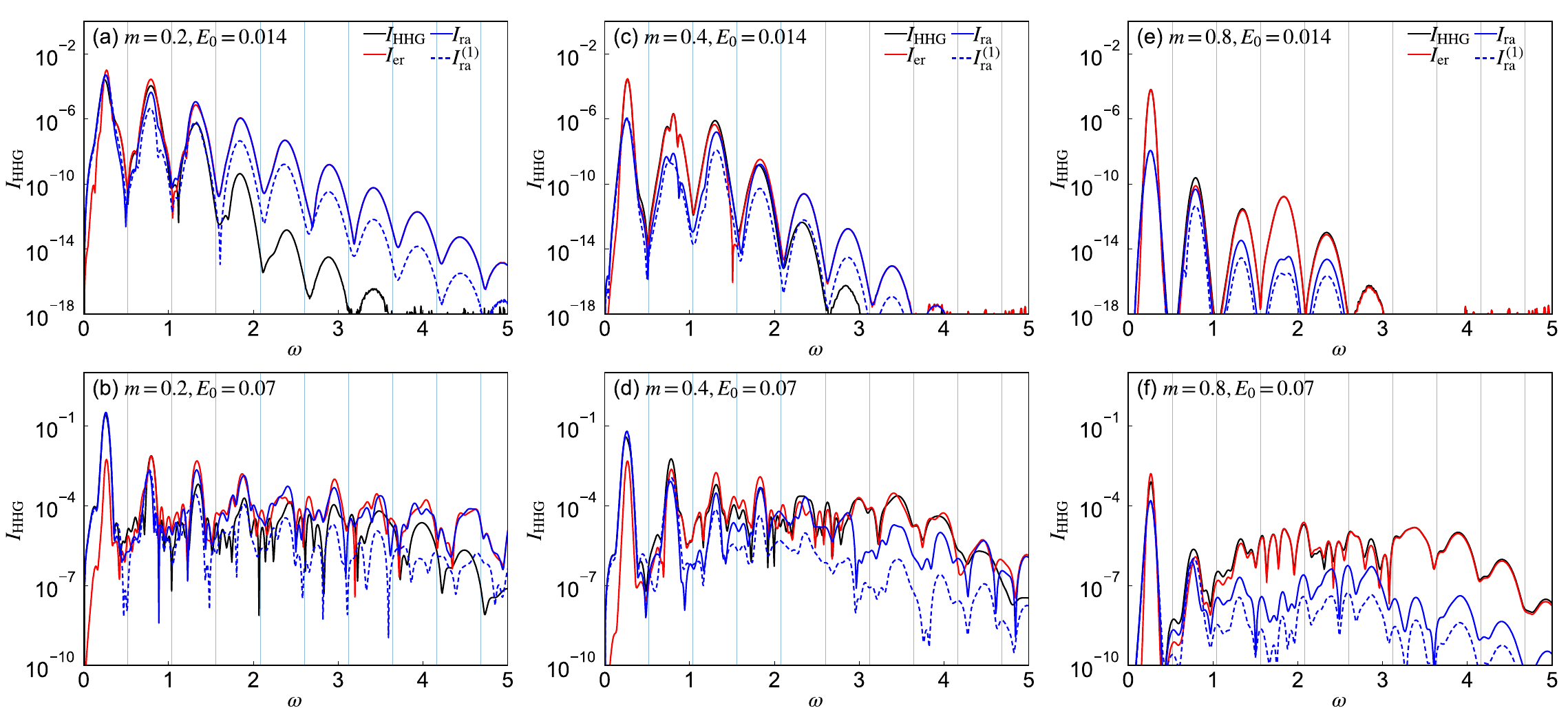} 
  \caption{(a-f) HHG spectra of gapped graphene for indicated values of the gap ($2m$) and field strength.  We compare contributions from different types of currents; 
  $I_{\rm HHG}=|\omega J_X(\omega)|^2$, $I_{\rm er}=|\omega J_{{\rm er},X}(\omega)|^2$, $I_{\rm ra}=|\omega J_{{\rm ra},X}(\omega)|^2$, $I^{(1)}_{\rm ra}=|\omega J^{(1)}_{{\rm ra},X}(\omega)|^2$.
In all cases, we use $t_{\rm hop}=3$, $\mu=0$, $T_1=150$ and $T_2=30$. The parameters of the electric field are $t_0=280,\sigma=40$ and $\Omega=0.26$. These results are obtained from the analysis of the Dirac models.}
  \label{fig:Dirac_HHG_mu_mass}
\end{figure*}

We consider the excitation with the linearly polarized light along the $X$ direction,
\eqq{
A_X(t) &= \frac{E_{0}}{\Omega} F_G(t,t_0,\sigma) \sin(\Omega (t-t_0)),
}
where $F_G(t,t_0,\sigma)=\exp[-\frac{(t-t_0)^2}{2\sigma^2}]$. We measure the HHG spectra polarized along the $X$ direction as $I_{\rm HHG}=|\omega J_X(\omega)|^2$. Here, $J_X$ indicates the current along the $X$ direction.
We note that due to the mirror symmetry along the $X$ direction (Fig.~\ref{fig:Graphene}), only odd harmonics are present in the HHG signal (finite even harmonics are due to the finite pulse used in the simulations).
Here, we focus on the system with vanishing gap ($m\rightarrow 0$), and study the doping dependence of HHG.
In Figs.~\ref{fig:Dirac_HHG_mu_dep}(a) and (b), we show how the HHG spectra change with modifying the chemical potential.
We also plot the intensity of the peaks in the HHG spectra ($I_{\rm peaks}$) as a function of the chemical potential in Figs.~\ref{fig:Dirac_HHG_mu_dep} (c) and (d).
Here, $I_{\rm peaks}(n)$ for the $n$th HHG peak is defined as $I_{\rm peaks}(n)=\int^{(n+\delta)\Omega}_{(n-\delta)\Omega} d\omega I(\omega)$, and we set $\delta=0.5$. When the field is relatively weak ($\simeq 1$ MV/cm), one can see the clear dependence of HHG on the chemical potential, where the HHG intensity 
can change by an order of magnitude.   In particular, the intensity of the 5th and 7th peaks increases with the doping from half filling, and the 5th peak intensity shows non-monotonic behavior.
The increase of the HHG intensity originates from that the cancellation between the intraband and interband current becomes less severe upon doping, as discussed below.
On the other hand, when the field is relatively strong ($\simeq 5$ MV/cm), the effects of doping become marginal. 
This change in the doping effects depending on the field strength can be understood by considering which electrons contribute to HHG.
In Figs.~\ref{fig:Dirac_HHG_mu_dep} (e) and (f), we show the intensity of the 5th harmonic peak of $I_{\rm HHG,\bk}(\omega) =|\omega J_{\bk,X} (\omega)|^2$. Behavior of the other harmonics is qualitatively the same. 
The results suggest that for the weaker field only the electrons around the Dirac points contribute to HHG, while for the stronger field the electrons in a larger range contribute to HHG.
This naturally explains the weak doping dependence of HHG for stronger fields, since the contribution from electrons around the Dirac point becomes less important.
In addition, Figs.~\ref{fig:Dirac_HHG_mu_dep} (e) and (f) tell that the contribution from electrons along the Dirac point is small.
This is natural since electrons along the Dirac point do not change the velocity under the field and thus do not contribute to HHG.  
Furthermore, the region of the strong contribution is extended along the field direction but limited in the perpendicular direction, suggesting that HHG mainly originates from the electrons moving along the optimal band dispersion.

Now, we study in detail the contributions from different types of currents and discuss the importance of $J_{\rm ra}^{(2)}$. 
In Figs.~\ref{fig:Dirac_HHG_detail}(a-d), we show the contributions from different types of currents; $I_{\rm HHG}=|\omega J_X(\omega)|^2$, $I_{\rm er}=|\omega J_{{\rm er},X}(\omega)|^2$, $I_{\rm ra}=|\omega J_{{\rm ra},X}(\omega)|^2$, $I^{(1)}_{\rm ra}=|\omega J^{(1)}_{{\rm ra},X}(\omega)|^2$.
For the first harmonics ($\omega\simeq \Omega$), in all cases, the agreement between $I_{\rm ra}$ and $I^{(1)}_{\rm ra}$ is good and the cancellation between $I_{\rm er}$ and $I_{\rm ra}$ is marginal.
On the other hand, one needs to pay attention for the higher harmonics.  
When the field is relatively weak ($\simeq 1$MV/cm) and $\mu=0$, $I_{\rm er}$ and $I_{\rm ra}$ take very close values and the total spectrum $I_{\rm tot}$ becomes 
much smaller than the former two. Namely, the contributions from $J_{\rm er}$ and $J_{\rm ra}$ cancel each other out. 
This is also the case for stronger fields [see Figs.~\ref{fig:Dirac_HHG_detail}(b) (d)], although the cancellation is less pronounced than in Fig.~\ref{fig:Dirac_HHG_detail}(a).
For these cases, the correct evaluation of $J_{\rm ra}$ is important. 
On the other hand, for the doped system and for relatively weak fields, the contribution from $J_{\rm er}$ is dominant for the 3rd, 5th and 7th harmonics [see Fig.~\ref{fig:Dirac_HHG_detail}(c)].
In this case, although the individual contributions $I_{\rm er}$ and $I_{\rm ra}$ are decreased away from half filling, the total HHG intensity can be enhanced, since the cancellation between them becomes less severe.
This explains the increase behavior of the peak intensity of the 5th and 7th harmonics shown in Fig.~\ref{fig:Dirac_HHG_mu_dep}(c).
As for the 3rd harmonics, the cancellation is not as severe as the higher harmonics even at half filling [see Fig.~\ref{fig:Dirac_HHG_detail}(a)], which makes the doping dependence different. 

When  $J_{\rm ra}(t)$ is evaluated without $J_{\rm ra}^{(2)}$, the contribution to HHG is underestimated [see $I_{\rm ra}^{(1)}$ in Fig.~\ref{fig:Dirac_HHG_detail}].
Then, the cancellation between  $J_{\rm er}$ and $J_{\rm ra}$  is underestimated and the HHG intensity is overestimated in general.
This can lead to qualitatively opposite prediction about the doping dependence of HHG:
when $J_{\rm ra}^{(2)}$ is not included, the HHG intensity decreases with the doping when the field is relatively weak [see Fig.~\ref{fig:Dirac_HHG_detail}(e)].
For stronger fields, the doping dependence becomes marginal, but the HHG intensity is strongly overestimated [see Fig.~\ref{fig:Dirac_HHG_detail}(f)].

 \begin{figure}[tb]
  \centering
    \hspace{-0.cm}
    \vspace{0.0cm}
\includegraphics[width=70mm]{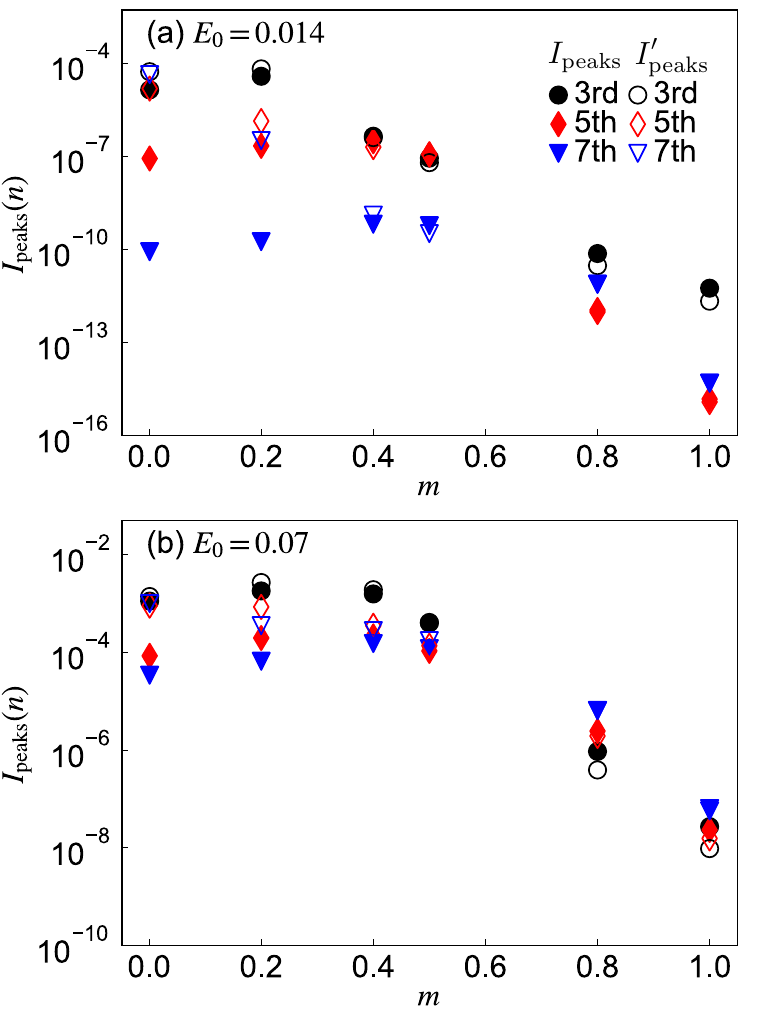} 
  \caption{The intensity of the peaks in the HHG spectra $I_{\rm peaks}(n)$, which is evaluated from $I_{\rm HHG}(\omega)$, and $I'_{\rm peaks}$, which is  evaluated from $I'_{\rm HHG}(\omega) =|\omega (J_{{\rm er},X}(\omega) + J_{{\rm ra},X}^{(1)}(\omega))|^2$ (i.e. without $J_{\rm ra}^{(2)}$), as a function of the mass term $m$. 
In all cases, we use $t_{\rm hop}=3$, $\mu=0$, $T_1=150$ and $T_2=30$. The parameters of the electric field are $t_0=280,\sigma=40$ and $\Omega=0.26$.  These results are obtained from the analysis of the Dirac models.}
  \label{fig:Dirac_HHG_m_dep}
\end{figure}

 \begin{figure}
  \centering
    \hspace{-0.cm}
    \vspace{0.0cm}
\includegraphics[width=85mm]{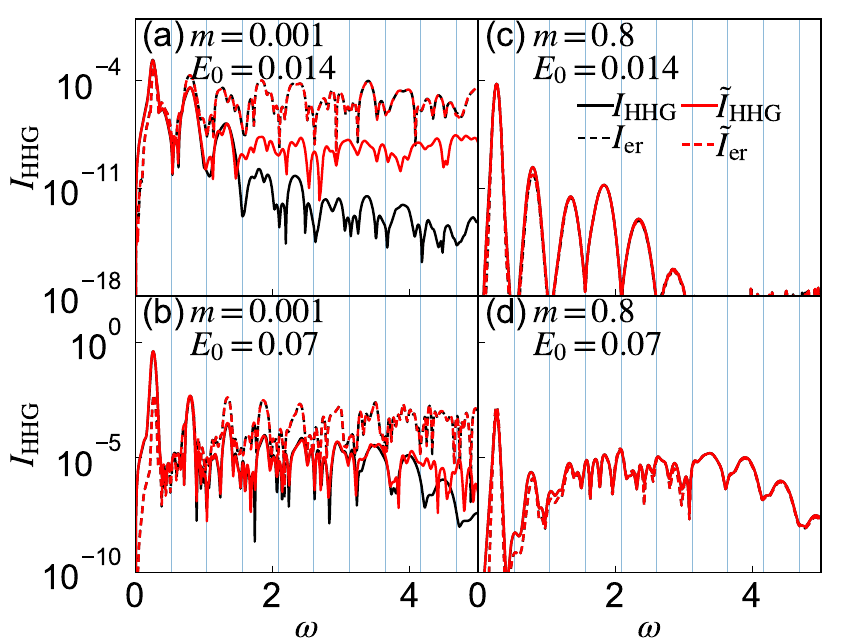} 
  \caption{ HHG spectra of gapped graphene for indicated values of the gap ($2m$) and field strength.  We compare contributions from different types of currents; 
  $I_{\rm HHG}=|\omega J_X(\omega)|^2$, $\tilde{I}_{\rm HHG}=|\omega (J_{{\rm ra},X}(\omega) -i\omega P_{{\rm er},X}(\omega))|^2$, $I_{\rm er}=|\omega J_{{\rm er},X}(\omega)|^2$ and $\tilde{I}_{\rm er}=|\omega^2 P_{{\rm er},X}(\omega)|^2$.
  In all cases, we use $t_{\rm hop}=3$, $\mu=0$,  $T_1=150$ and $T_2=30$. The parameters of the electric field are $t_0=280,\sigma=40$ and $\Omega=0.26$. These results are obtained from the analysis of the Dirac models.}
  \label{fig:Dirac_HHG_P_J}
\end{figure}

\subsection{Linearly polarized light: Effects of the mass term}
Now the question is when $J^{(2)}_{\rm ra}$ becomes important.
In order to obtain insight into this question, we examine the gap-size dependence of HHG at half filling [see Fig.~\ref{fig:Dirac_HHG_mu_mass}].
The results indicate that when the gap is small or comparable to the excitation frequency, contributions from the intraband and interband currents are comparable and cancel each other.
In this regime, the accurate evaluation of the intraband current is crucial, in order to correctly predict the dependence of HHG on system parameters.
On the other hand, when the band gap is sufficiently large compared to the excitation frequency, the contribution from the interband current becomes dominant for $\omega\gtrsim 2m$. 
Although there still remains substantial difference between the contributions from $J_{\rm ra}$ and $J^{(2)}_{\rm ra}$, the difference hardly affects the general structure of the HHG spectrum in this regime [see Figs.~\ref{fig:Dirac_HHG_mu_mass}(e,f)].

The cancellation between the intraband and interband currents mainly originates from $J^{(2)}_{{\rm ra},X}$ and $J^{(2)}_{{\rm er},X}(=-J^{(2)}_{{\rm ra},X})$.
As is indicated in Figs.~\ref{fig:Dirac_HHG_detail}(a)(b) and Figs.~\ref{fig:Dirac_HHG_mu_mass}(a)(b), when the gap is smaller than or comparable to the excitation frequency, these terms become the dominant components in the intraband current and the interband current, respectively.
Since ${\bm J}_{\rm ra}^{(2)}$ is the modulation of the intraband polarization by the interband transition, this term is expected to be large when the gap is small and 
the photo-excitation between the bands is activated. 
On the other hand, when the gap becomes larger, the contribution of these terms should be suppressed since there is no efficient transition by the photo-excitation,
and these terms should become less important.

To demonstrate the importance of the contribution of $J^{(2)}_{\rm ra}$, we show the gap-size dependence of the HHG peak in Fig.~\ref{fig:Dirac_HHG_m_dep}.
In the full evaluation, the peak intensity increases as the gap is increased from zero.
This can be understood by the fact that cancellation between the intraband and interband currents is relaxed.
On the other hand, when $J^{(2)}_{\rm ra}$ is not included, the intensity of the 5th and 7th harmonics is severely overestimated for small $m$
and the intensity is monotonically decreased. 
These results underpin the importance of the full evaluation of the current for small-gap systems.

Next we discuss the potential inconsistency between the different ways of evaluating the interband current [see Fig.~\ref{fig:Dirac_HHG_P_J}].
Within the present choice of $T_1$ and $T_2$, there is no crucial discrepancy between $I_{\rm er}$, which is evaluated from $\langle \hat{J}_{\rm er}\rangle$, and $\tilde{I}_{\rm er}$, which is evaluated from $\langle \hat{P}_{\rm er}\rangle$. We note that compared to the previous study~\cite{Wilhelm2021PRB}, which emphasizes the discrepancy between $\tilde{I}_{\rm er}$ and $I_{\rm er}$, we use much larger dephasing time.
However, there is clear difference between the full HHG spectra $I_{\rm HHG}$ and $\tilde{I}_{\rm HHG}(=|\omega (J_{\rm ra}(\omega) -i\omega P_{\rm er}(\omega))|^2)$ when the field is relatively weak and the mass is small [see Fig.~\ref{fig:Dirac_HHG_P_J}(a)].
This is natural since in this regime the cancellation between $I_{\rm er}$ and  $I_{\rm ra}$ is strong.
On the other hand, in the rest of the cases, where the cancellation is less severe, agreement between $I_{\rm HHG}$ and $\tilde{I}_{\rm HHG}$  becomes reasonable.

\subsection{Comments on circularly polarized light}
Finally, we comment on cases where we excite the system with the circularly polarized light;
\eqq{
A_X(t) &= \frac{E_{0x}}{\Omega}  F_G(t,t_0,\sigma) \cos(\Omega (t-t_0) -\frac{\pi}{4}) \\
A_Y(t) &= \frac{E_{0y}}{\Omega} F_G(t,t_0,\sigma) \cos(\Omega (t-t_0) + \frac{\pi}{4}).  \nonumber
}
We analyzed the HHG spectrum for various values of the ellipticity of the light.
However, we do not show the detailed results here, since the general tendency turns out to be essentially the same as the cases with the linearly polarized light.
Firstly, when the gap is small, the doping dependence of the HHG spectrum is smaller for cases with stronger laser fields, as in the cases with the linearly polarized light.
This is because the contributions from the electrons around the Dirac point becoming less important for stronger fields as in the cases with the linearly polarized field.
Secondly, the influence of $J_{\rm ra}^{(2)}$ also follows the same trend as the cases with the linearly polarized light.
When the gap is small, the cancellation between the contributions from $J_{\rm ra}$ and $J_{\rm er}$ is strong and full evaluation of $J_{\rm ra}$ is important.
On the other hand, when the gap becomes large compared to the excitation frequency, the contribution from  $J_{\rm er}$ becomes dominant and the contribution from $J_{\rm ra}^{(2)}$ becomes less relevant.
This feature can be explained from that contributions from $J^{(2)}_{\rm ra}$ and $J^{(2)}_{\rm er}$ should become large when the photo-excitation is activated for small gaps as in the cases with the linearly polarized field.
One of the important feature characteristic of HHG in solids is the dependence on the ellipticity~\cite{Yoshikawa2017Science,Sato2021PRB}. Namely, the HHG intensity can increase at nonzero ellipticity.
The discussion on the role of $J_{\rm ra}^{(2)}$  suggests that one needs to pay close attention when one evaluates the ellipticity-dependence of HHG for small gap systems like graphene.

\section{Conclusions}
In this paper, we studied the doping and gap-size dependence of HHG in gapped graphene under mid-infrared excitations and revealed the importance of a consistent representation of the light-matter coupling.
Focusing on the two-band systems, we explicitly revealed the relation between the frequently used representations of the SBEs, which are based on different gauges of the light and  bases for electric states.
As shown in Ref.~\onlinecite{Wilhelm2021PRB} for general cases, we pointed out several issues that may cause inconsistency between different representations.
In particular, we focus on the impact of a term in the intraband current $J_{\rm ra}^{(2)}$, which corresponds to the change of the intraband dipole via the interband transition and is often neglected in the HHG analysis.
With a systematic analysis of the doping and gap-size dependence of HHG in gapped graphene,
we showed that the contribution from $J_{\rm ra}^{(2)}$ is crucial when the gap is smaller than or comparable to the excitation frequency and that the evaluation without $J_{\rm ra}^{(2)}$ can lead to 
qualitatively opposite behavior of the dependence on parameters such as doping.
On the other hand, when the gap is large enough compared to the excitation frequency, the effects of $J_{\rm ra}^{(2)}$ are less important.
The theoretical insight into the  relation between frequently used representations and the importance of $J_{\rm ra}^{(2)}$ should be valuable to systematically understand how HHG changes with system parameters such as doping-level~\cite{Nishidome2020} and temperatures~\cite{Uchida2022PRL}.

In our study, we introduced the phenomenological relaxation/dephasing terms, and fixed their values. However, in practice, these values may change with doping-level~\cite{Nishidome2020} or with temperatures due to the correlation effects~\cite{Uchida2022PRL,Nagai2022,Du2022PRA,Murakami2022arxiv}.
Although recently the effects of correlations on HHG beyond the phenomenological description have been attracting much interest~\cite{Kemper2013NJP,Orlando2018,Silva2018NatPhoton,Murakami2018PRL,Murakami2018PRB,Markus2020,Tancogne-Dejean2018,Ishihara2020,Chinzei2020,Orthodoxou2021,Murakami2021PRB,Rostami2021PRB,Rostami2021,Can2022PRL,Hansen2022PRA,Murakami2022arxiv}, 
deeper understanding is required for further accurate understanding of behavior of HHG. This is an important future task.

\acknowledgements
We would like to acknowledge fruitful discussions with Kento Uchida, Hiroyuki Nishidome, Kohei Nagai, Kazuhiro Yanagi and Koichiro Tanaka.
This work is supported by Grant-in-Aid for Scientific Research from JSPS, KAKENHI Grant Nos. JP20K14412 (Y. M.), JP21H05017 (Y. M.), JST CREST Grant No. JPMJCR1901 (Y. M.).
M.S. thanks the Swiss National Science Foundation SNF for its support with an Ambizione grant (project No.~193527). 

\appendix
\section{Detail of the implementation}  \label{sec:Z}
In this paper, we study the tight-binding model and the Dirac models introduced in Secs.~\ref{sec:graphene_A} and \ref{sec:dirac} using the representation I, i.e. Eq.~\eqref{eq:vN_eq_T1T2_DW}.
In the practical implementation of  Eq.~\eqref{eq:vN_eq_T1T2_DW}, at each time step, we evaluate $ \brho^{\rm H}_\bk(t)$ using ${\bf U}^\dagger(\bk(t)) \brho^{\rm DW}_\bk(t) {\bf U}(\bk(t))$, extract the off-diagonal components and make an inverse transformation to evaluate the last term of Eq.~\eqref{eq:vN_eq_T1T2_DW}. We note that as far as $\bh(\bk(t)) \not \propto {\bf I}$ (${\bf I}$ is the identity matrix), this operation does not depend on the choice of the gauge of ${\bf U}(\bk(t))$.

In order to directly compare the results of the tight-binding model and the Dirac models (where a momentum cutoff $|\bk| < k_c$ has to be introduced),  
we evaluate observables by considering the difference from equilibrium $\delta \brho_{\bk}(t) = \brho_{\bk}(t) - \brho_{{\rm eq},\bk(t)}$, where $\bk(t)=\bk - q\bA(t)$.
Here, $\brho_{{\rm eq}}$ indicates the equilibrium SPDM. 
The value of physical quantities such as the energy and the current depend on the choice of $k_c$, but the deviation from the equilibrium hardly depends on this choice.
In practice, for the direct comparison of the HHG spectrum between the tight-binding model and the Dirac models, we evaluate the current using $\delta \brho_{\bk}(t)$, instead of $\brho_{\bk}(t)$,
for the Dirac models at ${\bm K}$ and ${\bm K}'$ and sum up these contributions [see Appendix. \ref{sec:B}].
We also evaluate the different types of currents from $\delta \rho_{\bk}(t)$. This procedure is justified by the fact that those currents are zero when they are evaluated from $\brho_{{\rm eq},\bk(t)}$.
Note that we need to be careful when the system is a topological state where the Chern number becomes nonzero and thus $J^{(1)}_{\rm ra}$ [see Eq.~\eqref{eq:J_ra_1}] can be nonzero even for $\brho_{{\rm eq},\bk(t)}$.

\section{Expression of the dipole moment for the Dirac model}  \label{sec:A}
For completeness, we show the expression of the dipole moment and its relevant quantities for the Dirac models~\eqref{eq:Dirac_1}.
We express the Dirac Hamiltonian as 
\eqq{
{\bm h}({\bk}) = B(\bk)
\begin{bmatrix}
\cos\theta_\bk & \sin\theta_\bk e^{i\phi_\bk}\\
 \sin\theta_\bk e^{-i\phi_\bk} & -\cos\theta_\bk
\end{bmatrix}}
with $ B(\bk)>0$. We consider the unitary matrix,
\eqq{
U(\bk) 
= 
\begin{bmatrix}
\cos\frac{\theta_\bk}{2}e^{i\frac{\phi_\bk}{2}} & -\sin\frac{\theta_\bk}{2}e^{i\frac{\phi_\bk}{2}} \\
\sin\frac{\theta_\bk}{2}e^{-i\frac{\phi_\bk}{2}} & \cos\frac{\theta_\bk}{2}e^{-i\frac{\phi_\bk}{2}} 
\end{bmatrix},\label{eq:gauge_type1}
}
which diagonalizes ${\bm h}({\bk})$ as 
$ U^\dagger (\bk) {\bm h}({\bk})U(\bk) = {\rm diag}[B_\bk,-B_\bk]$.
We consider the expression of the dipole moment for this transformation.

To be more explicit for the Hamiltonian \eqref{eq:Dirac_1} expanded around ${\bm K}$, we have 
\eqq{
&B_\bk =\sqrt{m^2 + t_{\rm hop,0}^2 (k_x^2 + k_y^2)} \nonumber \\
&\cos\theta_\bk = \frac{m}{B_\bk},\;\; \sin\theta_\bk = \frac{ |\tilde{F}(\bk)|}{B_\bk}\\
& \phi_\bk = \mathrm{arg}(-k_x - ik_y),\nonumber 
}
where $t_{\rm hop,0}\equiv 1.5t_{\rm hop}$ and $\tilde{F}(\bk) = t_{\rm hop,0} (k_x + ik_y) $.
By introducing $\kappa=\sqrt{k_x^2 + k_y^2}$ and $\gamma = t_{\rm hop,0}/m$,
the dipole moments are expressed as 
\eqq{
& d_{00,x}(\bk) = \frac{1}{2\sqrt{1+\gamma^2 \kappa^2}} \frac{k_y}{\kappa^2}, \;\;\; d_{00,y}(\bk) = -\frac{1}{2\sqrt{1+\gamma^2 \kappa^2}} \frac{k_x}{\kappa^2} \nonumber \\
& d_{11,x}(\bk) = -d_{00,x}(\bk), \;\;\;\;\; d_{11,y}(\bk) = -d_{00,y}(\bk) \nonumber \\
& d_{01,x}(\bk) = -\frac{\gamma}{2} \frac{1}{\kappa \sqrt{1+\gamma^2 \kappa^2}} \Bigl[ k_y + i \frac{k_x}{\sqrt{1+\gamma^2 \kappa^2}} \Bigl] \\
& d_{01,y}(\bk) = \frac{\gamma}{2} \frac{1}{\kappa \sqrt{1+\gamma^2 \kappa^2}} \Bigl[ k_x - i \frac{k_y}{\sqrt{1+\gamma^2 \kappa^2}} \Bigl] \nonumber \\
& d_{10,x}(\bk) =  d_{01,x}(\bk)^*, \;\;\;\;\;\;\; d_{10,y}(\bk) =  d_{01,y}(\bk)^* \nonumber
}

As for the Berry curvature ${\boldsymbol \Omega}_{n}(\bk)\equiv \nabla_\bk \times  \bd_{nn}(\bk)$, we only have the $z$ component,
which becomes
\eqq{
\Omega_{0,z}(\bk) = \frac{\gamma^2} {2 (\sqrt{1+\gamma^2 \kappa^2})^3 },  \ \Omega_{0,z}(\bk) = -\Omega_{1,z}(\bk).
}
For ${\br}_{nm,\alpha} (n\neq m)$, we find 
\begin{align*}
[\br_{01,x}]_x &= \gamma^3 \frac{k_xk_y}{\kappa(1+\gamma^2 \kappa^2)^{\frac{3}{2}}} + i\gamma^3  \frac{k_x^2}{\kappa(1+\gamma^2 \kappa^2)^{2}}  \nonumber \\
[\br_{01,x}]_y &= \frac{\gamma^3}{2} \frac{k_y^2-k_x^2}{\kappa(1+\gamma^2 \kappa^2)^{\frac{3}{2}}} + i\gamma^3  \frac{k_xk_y}{\kappa(1+\gamma^2 \kappa^2)^{2}}\nonumber  \\
[\br_{01,y}]_x &= \frac{\gamma^3}{2} \frac{k_y^2-k_x^2}{\kappa(1+\gamma^2 \kappa^2)^{\frac{3}{2}}} + i\gamma^3  \frac{k_xk_y}{\kappa(1+\gamma^2 \kappa^2)^{2}} = [\br_{01,x}]_y\nonumber \\
[\br_{01,y}]_y &=  -\gamma^3 \frac{k_xk_y}{\kappa(1+\gamma^2 \kappa^2)^{\frac{3}{2}}} + i\gamma^3  \frac{k_y^2}{\kappa(1+\gamma^2 \kappa^2)^{2}}  \\
[\br_{10,x}]_x &= [\br_{01,x}]_x^*, \;\;\;\;\; [\br_{10,x}]_y = [\br_{01,x}]_y^*\nonumber  \\
[\br_{10,y}]_x &= [\br_{01,y}]_x^*, \;\;\;\;\; [\br_{10,y}]_y = [\br_{01,y}]_y^* \nonumber 
\end{align*}

The Dirac Hamiltonian around ${\bm K}'$ (we denote the corresponding quantities by a bar, e.\, g. $\bar{\phi}_{\bm k}$) is closely related to the Hamiltonian expanded around ${\bm K}$. One finds
\eqq{
&\bar{\theta}_\bk = \theta_\bk,\;\;
\bar{\phi}_\bk = -\phi_\bk + \pi,\;\;
\bar{d}_{nn,a}(\bk) = -d_{nn,a}(\bk),  \nonumber \\
& \bar{d}_{01,a}(\bk) = -{\rm conj}(d_{01,a}(\bk)), \\
& \bar{\Omega}_{n,z}(\bk) = -\Omega_{n,z}(\bk),\;\; 
\bar{\br}_{nm,a} = -{\rm conj}(\br_{nm,a} ) \nonumber
}

 \begin{figure}
  \centering
    \hspace{-0.cm}
    \vspace{0.0cm}
\includegraphics[width=85mm]{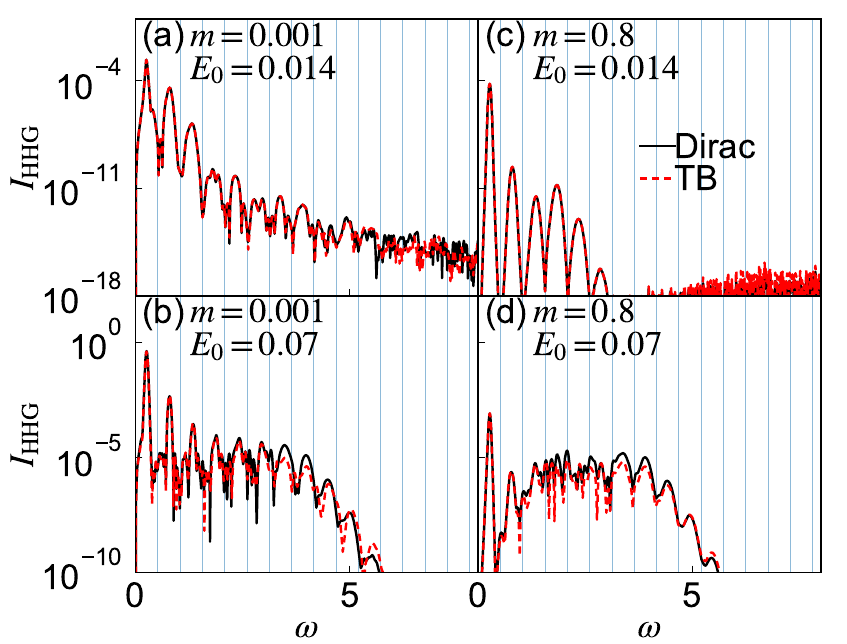} 
  \caption{Comparison of HHG spectra polarized along the $X$ direction evaluated with the tight-binding model or with the Dirac model for gapped graphene.  
  In all cases, we use $t_{\rm hop}=3$, $\mu=0$, $T_1=150$ and $T_2=30$. The parameters of the electric field are $t_0=280,\sigma=40$ and $\Omega=0.26$.}
  \label{fig:Graphene_VS_Dirac}
\end{figure}

\section{Tight-binding model vs Dirac model} \label{sec:B}
Here, we compare the HHG spectra polarized along the $X$ direction obtained by the analysis of the tight-binding model \eqref{eq:Graphene_TB_k} and that of the Dirac model \eqref{eq:Dirac_1} [see Fig.~\ref{fig:Graphene_VS_Dirac}].
The HHG spectra from the tight-binding model and the Dirac models match reasonably well for the excitation conditions used in this paper.
As we expected, agreement is better for the weaker field since the relevant electron dynamics is limited to the region around the Dirac point.
For the stronger field, agreement is better for the lower harmonics. This is also natural since the higher order harmonics involves the trajectory of electrons farther away from the Dirac points.


\bibliography{HHG_Ref}

\end{document}